\documentclass[draft]{agujournal2019}
\usepackage{url} %this package should fix any errors with URLs in refs.
\usepackage[inline]{trackchanges} %for better track changes. finalnew option will compile document with changes incorporated.
\usepackage{soul}
\draftfalse

\journalname{Geophysical Research Letters}

\begin{document}
%TC:ignore
%%%%%%%%%%%%%%%%%%%%%%%%%%%%%%%%%%%%%%%%%%%%%%%
%  TITLE
%
% (A title should be specific, informative, and brief. Use
% abbreviations only if they are defined in the abstract. Titles that
% start with general keywords then specific terms are optimized in
% searches)
%
%%%%%%%%%%%%%%%%%%%%%%%%%%%%%%%%%%%%%%%%%%%%%%%

% Example: \title{This is a test title}

\title{Lightning-Fast Convective Outlooks: Predicting Severe Convective Environments with Global AI-based Weather Models}
%Alternative titles: 
%\title{How well are Global AI-based Weather Models Predicting Severe Convective Environments?}
%\title{On the prediction of severe convective environments with global AI-based models}
%%%%%%%%%%%%%%%%%%%%%%%%%%%%%%%%%%%%%%%%%%%%%%%
%
%  AUTHORS AND AFFILIATIONS
%
%%%%%%%%%%%%%%%%%%%%%%%%%%%%%%%%%%%%%%%%%%%%%%%

% Authors are individuals who have significantly contributed to the
% research and preparation of the article. Group authors are allowed, if
% each author in the group is separately identified in an appendix.)

% List authors by first name or initial followed by last name and
% separated by commas. Use \affil{} to number affiliations, and
% \thanks{} for author notes.
% Additional author notes should be indicated with \thanks{} (for
% example, for current addresses).

% Example: \authors{A. B. Author\affil{1}\thanks{Current address, Antartica}, B. C. Author\affil{2,3}, and D. E.
% Author\affil{3,4}\thanks{Also funded by Monsanto.}}

\authors{Monika Feldmann\affil{1}%\thanks{monika.feldmann@unibe.ch}
, Tom Beucler\affil{2}, Milton Gomez\affil{2}, Olivia Martius\affil{1}}

 \affiliation{1}{Institute of Geography - Oeschger Centre for Climate Change Research, University of Bern}
 \affiliation{2}{Faculty of Geosciences and Environment - Expertise Center for Climate Extremes, University of Lausanne}

% \affiliation{3}{Third Affiliation}
% \affiliation{4}{Fourth Affiliation}

%\affiliation{=number=}{=Affiliation Address=}
%(repeat as many times as is necessary)

% Corresponding author mailing address and e-mail address:

% (include name and email addresses of the corresponding author.  More
% than one corresponding author is allowed in this LaTeX file and for
% publication; but only one corresponding author is allowed in our
% editorial system.)

% Example: \correspondingauthor{First and Last Name}{email@address.edu}

\correspondingauthor{Monika Feldmann}{monika.feldmann@unibe.ch}

%%%%%%%%%%%%%%%%%%%%%%%%%%%%%%%%%%%%%%%%%%%%%%%
% KEY POINTS
%%%%%%%%%%%%%%%%%%%%%%%%%%%%%%%%%%%%%%%%%%%%%%%
%  List up to three key points (at least one is required)
%  Key Points summarize the main points and conclusions of the article
%  Each must be 140 characters or fewer with no special characters or punctuation and must be complete sentences

% Example:
% \begin{keypoints}
% \item	List up to three key points (at least one is required)
% \item	Key Points summarize the main points and conclusions of the article
% \item	Each must be 140 characters or fewer with no special characters or punctuation and must be complete sentences
% \end{keypoints}

\begin{keypoints}
\item AI-based global weather models produce forecasts with sufficient accuracy to derive instability and shear metrics skillfully
\item The best AI-based weather models are capable of competing with state-of-the-art numerical weather predictions of instability and shear
\item This is a major step towards computationally inexpensive and fast convective outlooks
\end{keypoints}

%%%%%%%%%%%%%%%%%%%%%%%%%%%%%%%%%%%%%%%%%%%%%%%
%
%  ABSTRACT and PLAIN LANGUAGE SUMMARY
%
% A good Abstract will begin with a short description of the problem
% being addressed, briefly describe the new data or analyses, then
% briefly states the main conclusion(s) and how they are supported and
% uncertainties.

% The Plain Language Summary should be written for a broad audience,
% including journalists and the science-interested public, that will not have 
% a background in your field.
%
% A Plain Language Summary is required in GRL, JGR: Planets, JGR: Biogeosciences,
% JGR: Oceans, G-Cubed, Reviews of Geophysics, and JAMES.
% see http://sharingscience.agu.org/creating-plain-language-summary/)
%
%%%%%%%%%%%%%%%%%%%%%%%%%%%%%%%%%%%%%%%%%%%%%%%

%% \begin{abstract} starts the second page

% 147 / 150 words

%TC:endignore
\begin{abstract}
Severe convective storms are among the most dangerous weather phenomena and accurate forecasts mitigate their impacts. The recently released suite of AI-based weather models produces medium-range forecasts within seconds, with a skill similar to state-of-the-art operational forecasts for variables on single levels. However, predicting severe thunderstorm environments requires accurate combinations of dynamic and thermodynamic variables and the vertical structure of the atmosphere. Advancing the assessment of AI-models towards process-based evaluations lays the foundation for hazard-driven applications. We assess the forecast skill of three top-performing AI-models for convective parameters at lead-times of up to 10 days against reanalysis and ECMWF's operational numerical weather prediction model IFS. In a case study and seasonal analyses, we see the best performance by GraphCast and Pangu-Weather: these models match or even exceed the performance of IFS for instability and shear.  This opens opportunities for fast and inexpensive predictions of severe weather environments. 
\end{abstract}

%TC:ignore

\section*{Plain Language Summary}
%200 / 200 words
Over the past year, several global AI-based weather models were released and produce a similar quality of forecasts as traditional weather models. AI-models are very fast and computationally cheap to produce forecasts. The evaluation of AI-models has largely focused on single atmospheric variables at certain heights. To forecast specific phenomena, such as thunderstorms, a combination of variables must be accurate at multiple heights. Here we use the output of AI-models to derive thunderstorm-related parameters. We compare 10-day-forecasts between 3 AI-models and a state-of-the-art traditional weather model while using a reanalysis dataset as the reference.\\
With the example of a tornado outbreak in the southern United States, we see that all models are capable of forecasting thunderstorm parameters multiple days in advance. To obtain a robust assessment, we evaluate the entire thunderstorm season in 2020 in regions such as North America, Europe, Argentina, and Australia, where severe thunderstorms occur frequently. Two of the three AI-models achieve similar or even better results than the traditional weather model while being much cheaper to operate computationally.\\
Forecasting thunderstorm parameters directly, instead of calculating them afterwards, is likely to produce even better results. This opens opportunities for rapid and accessible forecasts for severe thunderstorm phenomena.

%%% CURRENTLY 3616 / 4000 words for the main body of the article + 300 words Fig captions - and 4 Figures %%%
%TC:endignore

\section{Introduction}

Artificial intelligence (AI)-based deterministic forecasts are revolutionizing the landscape of medium-range weather forecasting \cite{uphoff_aiweather_2023,bouallegue_aiforecast_2024}. Among the models released in the past year are FourCastNet \cite{bonev_fcn_2023}, GraphCast \cite{lam_graphcast_2023} and Pangu-Weather \cite{bi_pangu_2023}. They are capable of forecasting the main atmospheric state variables on various pressure levels within seconds \cite<on GPU,>[]{wong_deepmind_2023}, with a skill similar to the deterministic IFS model \cite<Integrated Forecasting System,>[operational version of IFS]{Bougeault_thorpex_2010} of the ECMWF (European Centre for Medium-range Weather Forecasts). AI-based weather predictions are rapidly revolutionizing weather forecasting \cite{bouallegue_aiforecast_2024,olivetti_aiextremes_2024,beucler_2024_nextgeneration,brunet_advancing_2023,Govett_exascale_2024,mcgovern_developing_2024} with scientists and forecasting centers calling for action to scrutinize AI-model performance from various perspectives, going beyond simply verifying the atmospheric state  \cite{uphoff_aiweather_2023,jeppesen_european_2024,bonavita_limitations_2024}. First studies investigate the robustness of forecasts for extremes \cite{charlton-perez_ai_2024,olivetti_extremes_2024} and model behavior in idealized experiments \cite{hakim_dyntest_2024}.\\
We focus on the specific, application-oriented task of accurately predicting atmospheric profiles in convective environments. The skillful prediction of convective environments requires high accuracy in the vertical profile and the simultaneous, correct prediction of both thermodynamic (temperature and humidity) and kinematic components (vertical wind shear). Convective storms require instability, governed by the vertical temperature and moisture profile \cite<e.g.,>[]{houze_clouddyn_2014,trapp_mesoscale_2013}. Severity is modulated by wind shear, emphasizing the importance of the wind profile \cite{houze_clouddyn_2014,trapp_mesoscale_2013}. By combining instability and wind shear, severe convective environments, that are related to particularly hazardous and impactful thunderstorms, can be identified \cite<e.g.,>[]{taszarek_severe_2020,kunz_convparam_2007,brooks_spatial_2003}.\\
Severe thunderstorms are among the most hazardous weather phenomena and caused substantial losses worldwide in recent years \cite<e.g.,>[]{Gallagher2024}. Severe thunderstorms can produce large hail and severe wind gusts, along with torrential rain, lightning strikes, and tornadoes \cite{houze_clouddyn_2014,Markowski_mesoscalemeteorology_2010}. \\
Global AI-models at medium-range lead-times currently provide weather information on a 0.25° grid. These models can hence not explicitly forecast severe storms. Therefore we focus on verifying pre-convective environments, characterized by mesoscale patterns of instability and wind shear. These environments provide information on the propensity of the atmosphere to produce severe storms \cite<e.g., >[]{taszarek_global_2021,taszarek_severe_2020,brooks_severe_2013} and are used to forecast severe convection \cite{battaglioli_hazardforecast_2023,brooks_estofex_2011,kunz_convparam_2007}. Accurate and inexpensive forecasts of mesoscale convective environments enable the timely assessment of severe convective outlooks for hazardous weather warnings.

\section{Data}
We focus our analysis on the year 2020, which is established as a reference year in the benchmark dataset ``Weatherbench2" \cite{rasp_weatherbench_2024} and was not used in training AI-models. We evaluate the three AI-models Pangu-Weather \cite{bi_pangu_2023}, GraphCast \cite{lam_graphcast_2023}, and FourCastNet \cite{bonev_fcn_2023}. We selected models with publicly accessible code and a stable version. Other AI-based global models have been developed \cite<AIFS, FuXi, Aurora and ClimaX;>[]{lang_aifs_2024,chen_fuxi_2023,bodnar_aurora_2024,nguyen_climax_2023}, but do not meet our evaluation requirements. The models are compared with the performance of the deterministic IFS-HRES \cite<>[IFS cycle 46r1]{ecmwf_ifs_2019}. The reference model IFS is the operational, physics-based, numerical weather prediction model at ECMWF, used for global medium-range forecasts \cite{thepaut_4dvar_1991,Bougeault_thorpex_2010}. The ERA-5 reanalysis \cite{era5} serves as a ground truth. We obtain Pangu-Weather, GraphCast, and IFS data through ``Weatherbench2" \cite{rasp_weatherbench_2024}. FourCastNet forecasts are produced with the ``ai-models'' code-release of ECMWF \cite{ai-models}.\\
In terms of architecture, Pangu-Weather is a transformer model that uses a 3-D visual earth transformer to encode the relative location on a sphere and additionally employs a hierarchical temporal aggregation, combining a 24h-prediction model with a 6h-prediction model \cite{bi_pangu_2023}. GraphCast is a graph neural network, which is particularly useful for spatially structured data, and uses successive 6h-predictions \cite{lam_graphcast_2023}. ``FourCastNet v2 small" is a transformer model, employing spherical Fourier neural operators (SFNO) to account for the spatial encoding of a sphere \cite{bonev_fcn_2023}.\\
All data is gridded at a 0.25° resolution, and IFS' native 0.125° resolution is regridded using the MIR method \cite<Meteorological Interpolation and Regridding,>[]{barratt_MIR_2017}. The forecasts are available twice daily at 00h UTC and 12h UTC up to 10 days lead-time in 6-hour increments. The AI-models are initialized from both ERA-5 and the IFS analysis. This reflects the conditions they were trained in (ERA-5) compared to an operational prediction set-up (IFS analysis). GraphCast has an additional operational version specifically tuned to being initialized with IFS data (GraphCast-oper). IFS is initialized with the IFS analysis.\\
For the computation of convective parameters, we use the temperature (T), specific humidity (Q), geopotential, and horizontal wind on the surface and pressure levels listed in Table S1 (see Section \ref{sec:cape}). The models differ in the available pressure levels and in the moisture variable that they predict, with most models providing 13 pressure levels between 50 and 1000 hPa and predicting Q. FourCastNet predicts relative humidity (RH), which needs to be converted to Q before deriving CAPE. An overview is provided in Table S1.

\section{Methods}

\subsection{Deriving convective parameters} \label{sec:cape}
We focus on the following convective parameters: convective available potential energy (CAPE), deep-layer shear (DLS), and a combined parameter of both called wmax-shear (WMS). CAPE is derived with the implementation for most unstable CAPE found in WRF-python \cite{wrf-python}, which launches a parcel profile above the surface at the height of highest equivalent potential temperature, using temperature, moisture, and geopotential at pressure levels, as well as surface elevation, pressure, and temperature. The relatively coarse vertical resolution reduces the accuracy of the CAPE calculation \cite{Wang_capeprec_2021,mensch_cape_2021}. However, we apply the same procedure to all datasets, hence treating them equally. We thereby artificially reduce the number of vertical levels available in the IFS and ERA-5 data. To approximate DLS, we compute the magnitude of the differential wind vector between the 10 m and the 500 hPa level. Further details on the computation of convective parameters are described in Text S1.

\subsection{Forecast evaluation}
To evaluate the forecast quality, we use several different scores. We first compute the root mean square error (RMSE) and the bias, two standard scores. To account for displacement errors and avoid the so-called `double penalty' \cite{Gilleland_verif_2009,Gilleland_prediction_2013}, we additionally use the fractional skill score \cite<FSS, >[]{mittermaier_fss_2021} and the structure, amplitude, and location (SAL) score, which evaluates these three properties individually \cite{wernli_salnovel_2008,wernli_spatial_2009}. Additional information on the interpretation of the scores is provided in Table S2.

We compute the FSS for two thresholds for each variable \cite<300 and 1000 J kg$^{-1}$ CAPE; 300 and 500 m$^{2}$ s$^{-2}$ WMS,>[]{taszarek_sounding-derived_2017}, with a neighborhood size of 1°, oriented on the gradients between categories of convective outlooks. The SAL is computed for 300 J kg$^{-1}$ CAPE, describing the convective environment. While the thresholds also include environments with only a moderate likelihood of convective storm formation, they serve as an estimated lower bound on severe convective storm occurrence.

\section{Case Study: US tornado outbreak}\label{sec:case}
\subsection{Synoptic situation}
\begin{figure}[htbp]
    \centering
    \includegraphics[width=.99\textwidth]{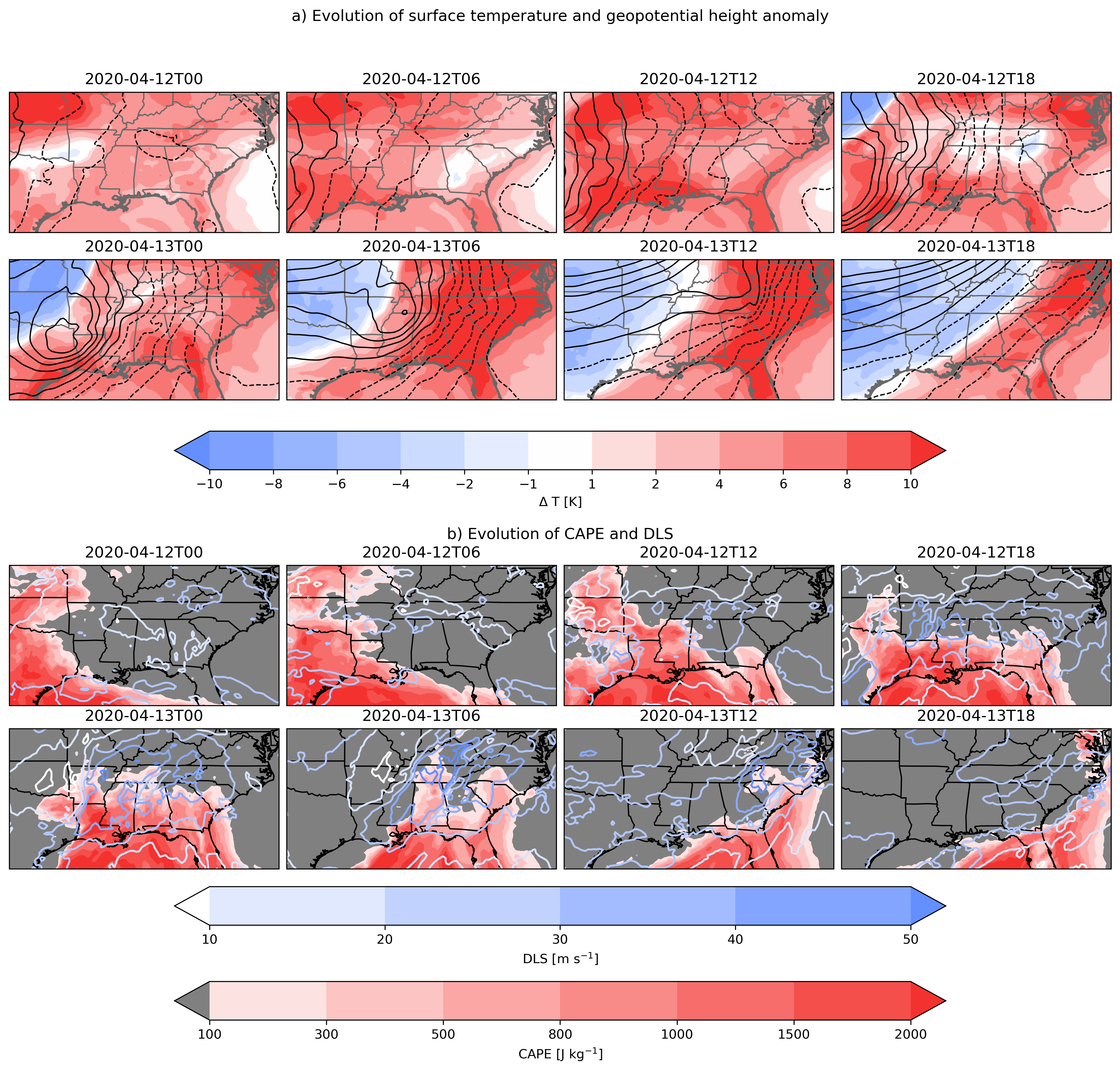}
    \caption{Synoptic situation in ERA-5 on April 12 and 13 2020, showing a deep trough moving across the United States, with anomalously warm temperatures and high CAPE values in the warm sector; a) Climatological geopotential height anomaly and surface temperature anomaly, contours indicate positive (dashed) and negative (solid) anomalies of geopotential height in 20 m increments b) pressure-level-derived CAPE and DLS, hatched area indicates where DLS \textgreater 20 m s$^{-1}$.}
    \label{fig:synop}
\end{figure}

Throughout 12 and 13 April 2020, 141 tornadoes were reported in 10 states, ranging up to a strength of EF4 \cite{spc_day1,spc_day2}. With an estimated 450 Million US\$ worth of property damage and 38 fatalities, this tornado outbreak was among the highest-impact events in the southern states \cite{whsv_impact_2020}, contributing to 2020 having the highest severe convective storm damages in the past \cite{swissre_2020}.\\
The tornado outbreak was characterized by an anomalously warm period in the preceding month \cite{copernicus_temp_2020}, leading to high sea surface temperatures in the Gulf of Mexico. The convective period was induced by a deep trough moving across the US. In the warm sector, warm and moist air from the Gulf was advected over the Southeastern states. Simultaneously, the approaching trough was associated with high DLS, which is crucial for convective organization. The convective potential of the situation was well forecast, the storm prediction center issued the first warnings four days ahead \cite{spc_warn_2020}.\\
While this event was most striking because of the large number of tornadoes, we focus on the convective environment characterized by CAPE and DLS. Figure \ref{fig:synop} shows the evolution of pressure-level CAPE, DLS, the 500 hPa geopotential height anomaly, and the surface temperature anomaly from ERA-5 during the outbreak.\\
The in-depth analysis of model forecast performance will focus on 12 April 2020 at 12:00 UTC, a few hours before the most intense storms occurred.

\subsection{Forecasting CAPE and DLS}
\begin{figure}[htbp]
    \centering
    \includegraphics[width=0.99\textwidth]{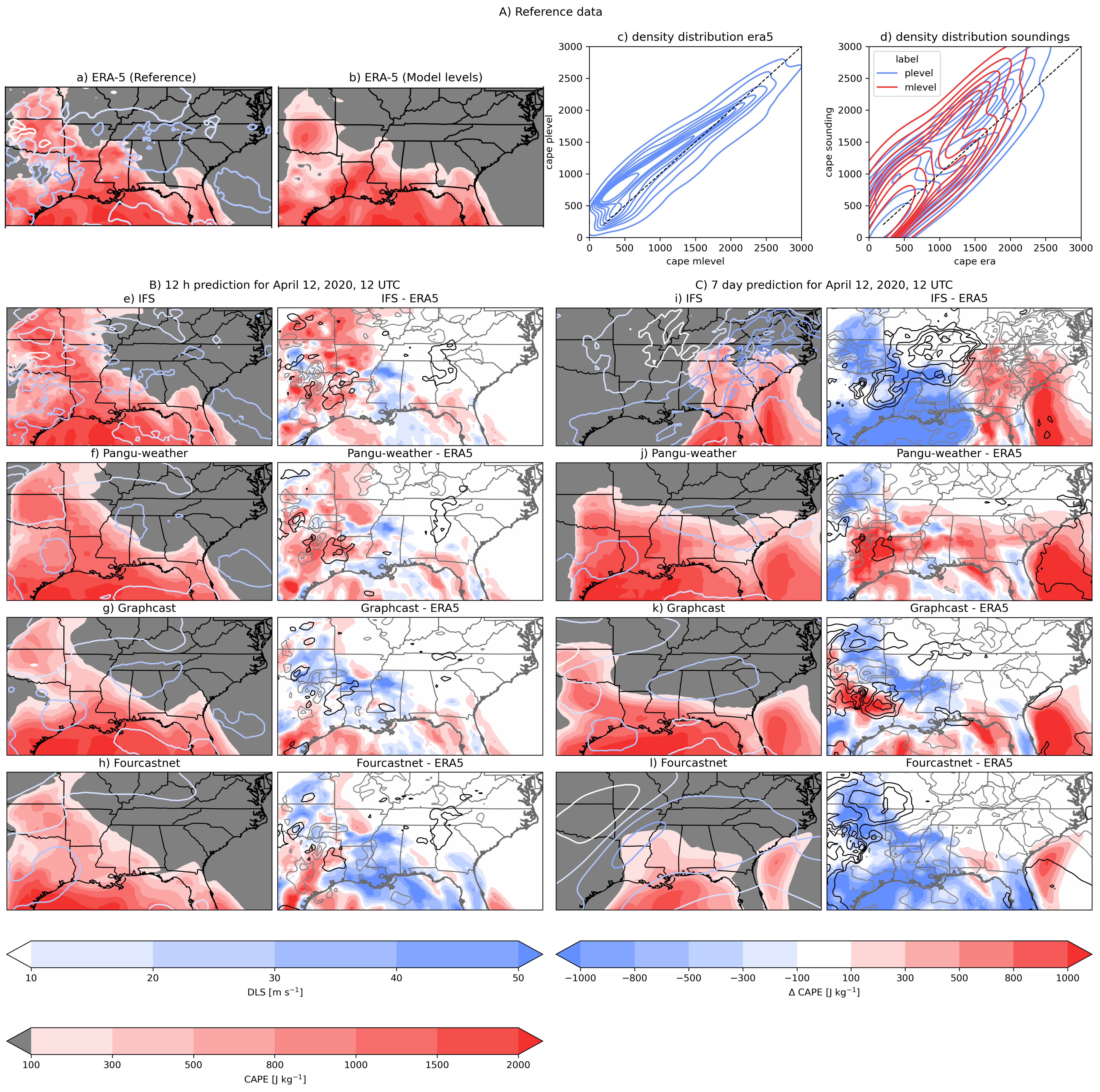}
    \caption{Forecast comparison of CAPE and DLS at 12h and 6.5 days lead-time; A) ERA-5 data of CAPE and DLS on April 12, 2020 at 12 UTC; a) CAPE and DLS from pressure levels, b) CAPE from model levels, c) density diagram of CAPE from pressure and model levels, d) density diagram of CAPE from pressure and model levels against observed soundings. A) 12h forecast of CAPE and DLS in comparison to ERA5 (e-h) B) 6.5-day day forecast of CAPE and DLS in comparison to ERA5 (i-l); contours indicated positive (gray) and negative (black) areas of $\Delta$ DLS in 5 m s$^{-1}$ increments}
    \label{fig:fc_comb}
\end{figure}

We established CAPE from ERA-5 pressure levels as the reference. However, to assess realistic convective outlooks, we additionally compare this to CAPE derived from model levels and soundings. Figure \ref{fig:fc_comb}A) shows a comparison of both (Fig. \ref{fig:fc_comb}a,b) and their density distribution over the entire case duration (Fig. \ref{fig:fc_comb}c). They correlate well (R$^2$=0.79), however, the distribution is narrowed, with an overestimation of low values and an underestimation of high values. When comparing to soundings (Fig. \ref{fig:fc_comb}d), any ERA-5 derived CAPE generally tends to be underestimated with respect to the radio-sounding. Diagrams of the observed soundings and ERA-5-modeled pseudo-soundings are provided in Fig. S1.\\
We first determine whether AI-models overall produce realistic CAPE and DLS values at a short lead-time of 12 hours. Fig. \ref{fig:fc_comb}B) shows all models initialized with IFS analysis (Fig. \ref{fig:fc_comb}e-h, absolute and difference to ERA-5). All models produce CAPE values in the right order of magnitude with a spatial distribution of DLS that corresponds well to the reference. Nonetheless, we can see differences between the models. All AI-model fields appear smoother than the reference data. IFS has the highest degree of spatial detail in both DLS and CAPE, while the AI-models mostly capture the larger-scale features. The high degree of detail in IFS, however, is penalized in the standard performance scores, as already a small displacement of features can lead to increases in RMSE (Fig. \ref{fig:fc_comb}e, RMSE-IFS=312 J kg$^{-1}$, RMSE-GraphCast-oper=271 J kg$^{-1}$; for all scores see Table S4). FourCastNet-oper underestimates CAPE in this case (Fig. \ref{fig:fc_comb}h). Of the AI-models, GraphCast-oper shows the closest match in terms of location and magnitude to ERA-5 (Fig. \ref{fig:fc_comb}h). \\

We next inspect the lead-time 156 hours / 6.5 days (Fig. \ref{fig:fc_comb}C). %Once again, IFS has a high degree of spatial detail, whereas the AI-models are smooth. Their smoothness does not appear much more pronounced than in the 30-hour forecast which is reflected in a marginally increased structure score (see Tables S4 and S5). 
At this longer lead-time, IFS displaces the event eastwards (Fig. \ref{fig:fc_comb}i), however, the magnitude of CAPE values is correct. This leads to large absolute errors (RMSE-IFS 829 J kg$^{-1}$). Pangu-Weather-oper's forecast also suffers from displacement error, owed to a very different spatial structure of the event (Fig. \ref{fig:fc_comb}j, RMSE-PanguWeather-oper 570 J kg$^{-1}$). %While the coastal areas still have high CAPE values the spatial distribution of CAPE is not reproduced well. 
GraphCast-oper has the best agreement in terms of event structure and location (Fig. \ref{fig:fc_comb}k, RMSE-GraphCast-oper 520 J kg$^{-1}$). FourCastNet-oper strongly underestimates CAPE and also has the largest degree of smoothing (Fig. \ref{fig:fc_comb}l). It misses most land areas that experience high CAPE and DLS. Nonetheless, we highlight here, that all models capture an area of elevated CAPE values more than 6 days in advance, in co-location with high DLS.\\

\section{Seasonal and regional analyses}\label{sec:season}
To investigate performance scores, we derive CAPE and WMS for North America, Europe, Argentina, and Australia and their respective convective seasons. CAPE and WMS are filtered by a land-sea-mask to focus on convectively relevant areas. Table S3 summarizes the regions' extent and seasons. We then compute the lead-time-dependent scores for each model.

\subsection{The North American convective season}
\begin{figure}[htbp]
    \centering
    \includegraphics[width=.99\textwidth]{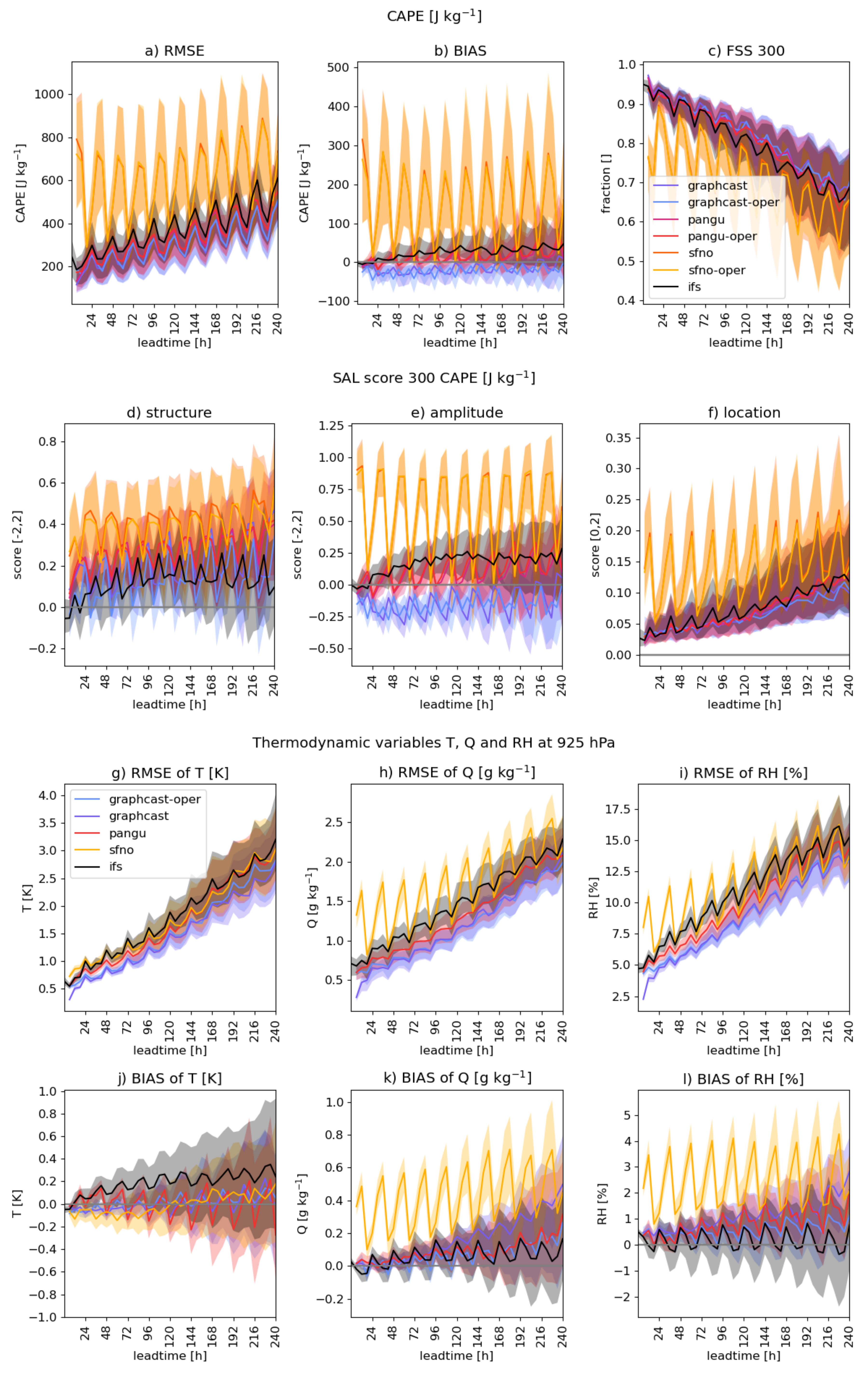}
    \caption{Seasonal evaluation of CAPE for North America at 00 UTC initialization shows Pangu-Weather and GraphCast outperforming IFS, with a moisture bias impacting the performance of FourCastNet. Top row: a) RMSE, b) BIAS, c) FSS$_{300}$; Second row: SAL components for CAPE \textgreater 300 J kg$^{-1}$; color legend in panel c); Third row: RMSE of T, Q and RH at 925 hPa (g-i); Fourth row:  BIAS (j-l); model legend in panel g). Solid line depicts the median and shading the interquartile range.}
    \label{fig:NA_scores}
\end{figure}

Figure \ref{fig:NA_scores} depicts the lead-time-dependent scores throughout the entire convective season for the USA. We include here both the models initialized with ERA-5 and with IFS analysis, initialized at 00 UTC.

The RMSE of CAPE (Fig. \ref{fig:NA_scores}a) shows close proximity of the scores of Pangu-Weather, GraphCast, and IFS. They lie well within each other's interquartile range (IQR). FourCastNet is a notable exception, performing consistently worse than all other models. GraphCast and Pangu-Weather tend to outperform IFS slightly. Since they are optimized on RMSE (albeit on different variables), we expect them to perform best in this metric. The BIAS (Fig. \ref{fig:NA_scores}b) shows more diversity, with IFS having an increasingly positive BIAS and Pangu-Weather a quite stationary BIAS, centered on 0. %Pangu-Weather's BIAS has fluctuations on a 24h-cycle owed to its integration structure using two different models. 
GraphCast moderately underestimates CAPE, while %, which is much improved in the retuned Graphcast-oper.
FourCastNet strongly overestimates CAPE, contrasting with the underestimation in the case study. The seasonal analysis is dominated by the principal convective season in the Great Plains, whereas the case study focuses on a spring convective event in the Southeast. For the FSS$_{300}$ (Fig. \ref{fig:NA_scores}c), GraphCast and Pangu-Weather are slightly better than IFS at intermediate lead-times. FourCastNet is consistently worse, largely lying outside of the IQR of the other models. In terms of SAL scores, the amplitude score mirrors the BIAS (Fig. \ref{fig:NA_scores}b,e). The positive structure score shows the smoother behavior of the AI-models (Fig. \ref{fig:NA_scores}d). It is important to note, that structure, amplitude, and location are not wholly independent of one another, as strong biases also affect structure and location of objects. %The overall bad performance of FourCastNet impacts the SAL score's ability to accurately characterize its field. %Most notably, the strongly increased location score is evident (Fig. \ref{fig:NA_scores}f), which may indicate that the overall structures of the CAPE field \textgreater 300 J kg$^{-1}$ are too different to appropriately compare to ERA5. 
%The other models behave very similarly regarding the location score and perform much better than FourCastNet. 
The location score shows the largest dependence on lead-time, steadily increasing (Fig. \ref{fig:NA_scores}f). Overall IFS scores close to `0' in both structure and amplitude, while the location score shows the strongest deterioration over lead-time.\\
A clear pattern visible in all models is a regular spikiness on a 24-hour frequency, when separating the data into the 00 UTC (Fig. \ref{fig:NA_scores}) and 12 UTC (SI Fig. S2) initializations. This is owed to the strong diurnal cycle of CAPE, which depends on the diurnal cycle of temperature \cite{leon_addressing_2018,lawson_diurnal_2016} and humidity. The models tend to perform better at night and in the early morning when CAPE is naturally low. The errors are higher during the day and in the evening when peak CAPE values occur. For Pangu-Weather this additionally coincides with the change between the 6h- and 24h-integration models, leading to a 24h cycle of skill even in combined initialization times (see Fig. \ref{fig:cape_reg}).
%Given that the initializations of all models are twice a day, regular variation in e.g., BIAS can be attributed to the strong diurnal cycle of CAPE, which depends on the diurnal cycle of temperature \cite{leon_addressing_2018,lawson_diurnal_2016}. By separating the data into the 00 UTC and 12 UTC initializations, a clear diurnal cycle emerges (figure not shown). The models tend to perform better at night and in the early morning when CAPE is naturally low. The errors are higher during the day and in the evening when peak CAPE values occur.\\ 

Given the similarity between the initializations with ERA5 and IFS analysis, we focus the next analyses on the initializations with IFS (previously flagged as ``oper"), mimicking an operational context. Since GraphCast (initialized with ERA-5) and GraphCast-oper (initialized with IFS) are two different model versions, we maintain both.

\subsection{The impact of thermodynamic conversions}\label{sec:tdyn}
With FourCastNet being an outlier, one core difference is the prediction of RH instead of Q. Q is used most commonly, but due to the relatively strong boundedness, RH has been shown to generalize better in some machine learning applications \cite{beucler_climate-invariant_2024,lin_stresstesting_2024}. This mostly applies to large domain shifts under climate change, where Q is no longer well-constrained. At the weather timescale, Q can be assumed to be in a stable value range. To calculate CAPE, RH needs to be converted to Q. The conversion is dependent on the temperature and pressure, making this an additional processing step, where forecast errors can compound. To quantify this effect, we additionally compare all models' RMSE and BIAS for RH, Q, and T in Figure \ref{fig:NA_scores}.\\

The RMSE (Fig. \ref{fig:NA_scores}g) and BIAS (Fig. \ref{fig:NA_scores}j) of T are quite similar across all models, indicating that the large differences in CAPE are not primarily owed to T. At short lead-times FourCastNet performs only marginally worse. Looking at Q, it becomes evident that FourCastNet not only has a larger RMSE (Fig. \ref{fig:NA_scores}h) but also consistently overestimates Q (Fig. \ref{fig:NA_scores}k), which does not improve much at short lead-times. For RH it becomes evident, that despite this being the natively predicted variable, FourCastNet also does not perform particularly well. The BIAS (Fig. \ref{fig:NA_scores}l) remains consistently positive. GraphCast-oper and Pangu-Weather consistently have a lower RMSE and BIAS (Fig. \ref{fig:NA_scores}i,l), despite RH being a converted variable in their output. Consequently, FourCastNet suffers from a double penalty - the moderately good RH undergoes further conversion to Q before being integrated into the computation of CAPE. Moisture is a key component in determining the lifting condensation level, which strongly impacts the integral of CAPE, leading to large compounding errors.

\subsection{Global convective hotspots}

\begin{figure}[htbp]
    \centering
    \includegraphics[width=0.99\textwidth]{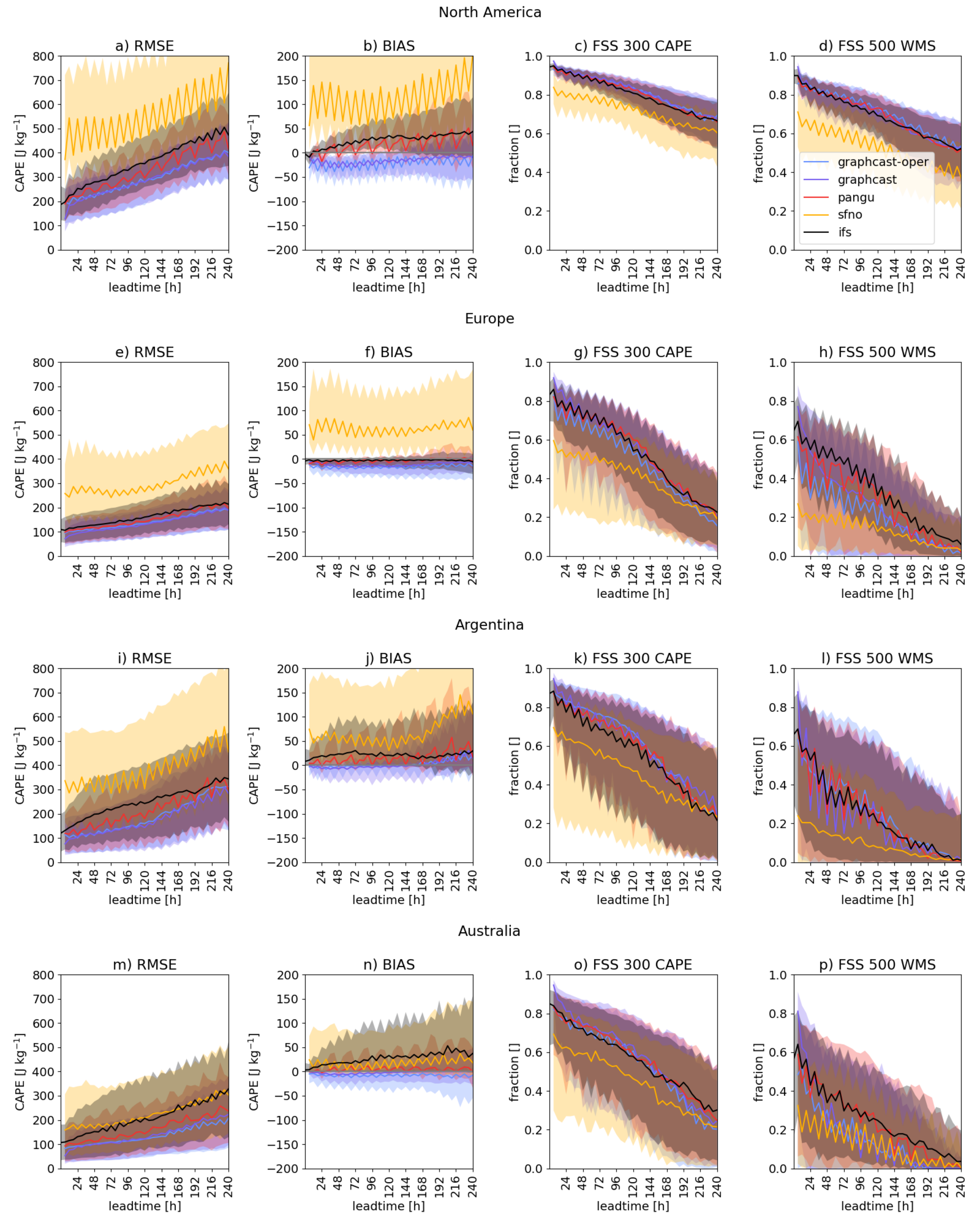}
    \caption{Seasonal evaluation of CAPE in North America (Apr-Aug, a-d), Europe (Apr-Sep, e-h), Argentina (Sep-Feb, i-l) and Australia (Sep-Feb, m-p) shows good performance of GraphCast and Pangu-Weather; solid line depicts median of score, the IQR is shown in shading}
    \label{fig:cape_reg}
\end{figure}

We next compare the models' performance in additional convectively active regions: Europe, Argentina, and Australia. Figure \ref{fig:cape_reg} depicts the RMSE, BIAS, and FSS$_{300}$ for CAPE, and the FSS$_{300}$ for WMS, computed on both the 00 UTC and 12 UTC initializations.

At first glance, the RMSE is highest in North America and Argentina (Fig. \ref{fig:cape_reg}a and i). This is partly due to the overall magnitude of CAPE values during the 2020 convective seasons; a value of 300 J kg$^{-1}$ falls at the 84th and 86th percentiles of the CAPE distribution in these regions, compared to the 90th and 93rd percentiles in Australia and Europe, respectively. The magnitude of CAPE impacts the FSS$_{300}$, as the FSS generally decreases as the threshold approaches the tail end of the overall value distribution. Hence it is consistent that the FSS$_{300}$ is overall best in North America (Fig. \ref{fig:cape_reg}c).\\
FourCastNet generally performs worse and has a wider range of performance scores than the other models. This is primarily driven by the quality of its RH forecast. Australia is an exception, where FourCastNet achieves similar RMSE (Fig. \ref{fig:cape_reg}m) to the other models, albeit with a very large IQR. The lower RMSE of CAPE stems from a better prediction of RH in this region (figure not shown).\\
In all regions, GraphCast-oper and GraphCast show the lowest RMSE of CAPE (Fig. \ref{fig:cape_reg}a,e,i,m) and over most lead-times the highest FSS$_{300}$ (Fig. \ref{fig:cape_reg}c,g,k,o), followed by Pangu-Weather and IFS. Despite the small differences, the high consistency across multiple regions shows this as systematic behavior.\\
The FSS$_{300}$ of CAPE and FSS$_{500}$ of WMS show similar behavior, with the WMS achieving lower scores, being at a higher convective threshold. This indicates that skill is primarily driven by CAPE, while shear plays a minor role in the predictability of convective environments. Both FSS show very similar performance for GraphCast, Pangu-Weather, and IFS. There is no consistent ranking as in the RMSE, highlighting the importance of considering multiple performance scores.

%To evaluate the co-occurrence of CAPE and DLS, an additional analysis of WMS can be found in Figure S2. The overall performance of WMS is dominated by CAPE, which is more difficult to predict than DLS.

\section{Discussion and Conclusion}

As shown in Section \ref{sec:season}, Pangu-Weather, GraphCast-oper, and GraphCast are capable of producing realistic forecasts of the mesoscale convective environment at medium-range lead-times in a matter of minutes and perform similarly to IFS. FourCastNet has the strongest deficiencies, partially owed to the additional processing step from RH to Q. However, as Section \ref{sec:tdyn} shows, forecasting errors in convective parameters are also owed to overall issues in forecasting moisture. With GraphCast and Pangu-Weather providing reasonable RH values derived from Q, we conclude that RH is not necessarily a worse choice than Q, but rather that ``FourCastNet v2 small" specifically has issues in forecasting moisture. The best choice in moisture variable depends on the application, as every thermodynamic conversion in a non-physical model comes with additional uncertainty.\\

The evaluation of different regions shows the global capability of Pangu-Weather and GraphCast. Despite regional performance differences, the ranking of models remains stable, with GraphCast consistently performing best in terms of RMSE, followed by Pangu-Weather, and IFS. These models score very similarly, and their systematic differences are smaller than the uncertainty of the initialization. For the FSS, Pangu-Weather and GraphCast are able to achieve similar scores to IFS, but do not consistently outperform it. This highlights how the RMSE, which is used in model training, is an incomplete perspective on model performance.\\

The perceived discrepancy between the case study and the evaluation of the North American convective season shows how case studies can be misleading for performance evaluation. The case study takes place in spring in the Southeastern States, while the bulk of the convective season is driven by the Great Plains in summer. Particularly the moderately good performance of FourCastNet in this instance is not typical when over most of the summer season, RH and subsequently CAPE are largely overestimated. An additional example case during the summer is provided in Figure S2. This highlights the necessity to evaluate a larger sample of cases, before drawing conclusions on AI-model-performance. While a single convective season is still rather short, the consistency of the results across four different global regions indicates robustness in our evaluation.\\

Despite the shortcomings of deriving CAPE from vertically coarse pressure levels, the promising results of GraphCast and Pangu-Weather indicate that CAPE is a worthwhile prediction variable. Predicting CAPE directly by training on ERA-5's model-level CAPE instead of deriving it, would likely be the most skillful approach. For instance, foundation models are designed for easier addition of variables after the main training of the model.\\
While forecasting convective environments is a large step closer to application-oriented forecasting, it still lacks the specificity of impact-oriented hazard predictions. Rapid progress in downscaling through super-resolution may enable direct hazard forecasts in the foreseeable future.\\
Currently, the vertical coarseness limits the number of convectively relevant variables that can be derived. We refrained from evaluating convective inhibition, near-surface shear, or helicity due to the limited resolution. However, these are key metrics for assessing convective initiation and convective hazards, especially tornadoes. With forecasters exploiting a myriad of convective parameters, it remains to be seen how practicable the direct prediction of additional convective parameters is in contrast to increasing the vertical resolution.\\
Taking this first step of deriving CAPE and DLS in AI-models already provides useful information for severe convective outlooks, which are key for severe weather warnings. This highlights how application-oriented products can be derived from AI-generated forecasts, with similar skill to operational numerical models. Especially with their very short runtime, they can deliver very timely forecasts at a low computational cost. This step towards process-based assessment of AI-models is crucial for the future development of hazard-driven applications.
%TC:ignore

%\section*{Author contributions}

%Text here ===>>>

%%%%%%%%%%%%%%%%%%%%%%%%%%%%%%%%%%%%%%%%%%%%%%%
%
% DATA SECTION and ACKNOWLEDGMENTS
%
%%%%%%%%%%%%%%%%%%%%%%%%%%%%%%%%%%%%%%%%%%%%%%%

\section*{Open Research Section}
The IFS, ERA-5, Pangu-Weather, and GraphCast data were accessed through ``Weatherbench2" \cite{rasp_weatherbench_2024}. FourCastNet predictions were computed with the code release of ECMWF \cite{ai-models}. The code used to process the forecast data and produce the visualizations can be accessed at \url{https://github.com/feldmann-m/AI-storm} with the assigned DOI \url{https://doi.org/10.5281/zenodo.12527391}.

\section*{Competing interests}
MF and OM are in positions funded by the Mobiliar Insurance Group. This funding source played no role in any part of the study.

%TC:endignore
\acknowledgments
We thank Zied Ben Bouallegue and Matthew Chantry for insightful discussions on the current state of AI-weather-forecasting. We also thank Louis Poulain-Auzéau for his work on installing the AI-models.\\
OM acknowledges support from the Swiss National Science Foundation project No. CRSII5\_201792.\\
MF processed the data, created the visualizations, and wrote the manuscript. TB provided computational resources and expertise on the properties of AI-models and their evaluation. MG produced the FourCastNet predictions, assisted with computational aspects, and provided valuable expertise. OM co-designed the project and advised it. All authors reviewed the manuscript.

\bibliography{references}

%Reference citation instructions and examples:
%
% Please use ONLY \cite and \citeA for reference citations.
% \cite for parenthetical references
% ...as shown in recent studies (Simpson et al., 2019)
% \citeA for in-text citations
% ...Simpson et al. (2019) have shown...
%
%
%...as shown by \citeA{jskilby}.
%...as shown by \citeA{lewin76}, \citeA{carson86}, \citeA{bartoldy02}, and \citeA{rinaldi03}.
%...has been shown \cite{jskilbye}.
%...has been shown \cite{lewin76,carson86,bartoldy02,rinaldi03}.
%... \cite <i.e.>[]{lewin76,carson86,bartoldy02,rinaldi03}.
%...has been shown by \cite <e.g.,>[and others]{lewin76}.
%
% apacite uses < > for prenotes and [ ] for postnotes
% DO NOT use other cite commands (e.g., \citet, \citep, \citeyear, \nocite, \citealp, etc.).
%

\end{document}

%% ------------------------------------------------------------------------ %%
%
%  TITLE
%
%% ------------------------------------------------------------------------ %%

%\includegraphics{agu_pubart-white_reduced.eps}

\title{Supporting Information for ``Lightning-Fast Convective Outlooks: Predicting Severe Convective Environments with Global AI-based Weather Models''}
%
% e.g., \title{Supporting Information for "Terrestrial ring current:
% Origin, formation, and decay $\alpha\beta\Gamma\Delta$"}
%
%DOI: 10.1002/%insert paper number here%

%% ------------------------------------------------------------------------ %%
%
%  AUTHORS AND AFFILIATIONS
%
%% ------------------------------------------------------------------------ %%

% List authors by first name or initial followed by last name and
% separated by commas. Use \affil{} to number affiliations, and
% \thanks{} for author notes.
% Additional author notes should be indicated with \thanks{} (for
% example, for current addresses).

% Example: \authors{A. B. Author\affil{1}\thanks{Current address, Antartica}, B. C. Author\affil{2,3}, and D. E.
% Author\affil{3,4}\thanks{Also funded by Monsanto.}}

\authors{Monika Feldmann\affil{1}, Tom Beucler\affil{2}, Milton Gomez\affil{2}, Olivia Martius\affil{1}}

% \affiliation{1}{First Affiliation}
% \affiliation{2}{Second Affiliation}
% \affiliation{3}{Third Affiliation}
% \affiliation{4}{Fourth Affiliation}

 \affiliation{1}{Institute of Geography - Oeschger Centre for Climate Change Research, University of Bern}
 \affiliation{2}{Faculty of Geosciences and Environment - Expertise Center for Climate Extremes, University of Lausanne}
%(repeat as many times as is necessary)
%\correspondingauthor{Monika Feldmann}{monika.feldmann@unibe.ch}

%% ------------------------------------------------------------------------ %%
%
%  BEGIN ARTICLE
%
%% ------------------------------------------------------------------------ %%

% The body of the article must start with a \begin{article} command
%
% \end{article} must follow the references section, before the figures
%  and tables.

\begin{article}

%\tableofcontents
\noindent\textbf{Contents of this file}

\begin{enumerate}
\item Text S1: Computation of convective parameters
\item Table S1: Model Information
\item Table S2: Performance Scores
\item Table S3: Convective Regions
\item Table S4: Forecast scores of April 13, 06 UTC, 30h lead-time
\item Table S5: Forecast scores of April 13, 06 UTC, 74 lead-time
\item Figure S1: Comparison of vertical profiles for April 12, 2020, 12 UTC
\item Figure S2: 12 UTC initializations for seasonal scores in the USA
\item Figure S3: CAPE and DLS forecasts for July 7, 2020, 06 UTC
%\item Figure S4: Skill of compounding CAPE and DLS
\end{enumerate}

\noindent\textbf{Introduction}
%Type or paste your text here. The introduction gives a brief overview of the supporting information. You should include information %about as many of the following as possible (when appropriate):
% 1. a general overview of the kind of data files;
% 2. information about when and how the data were collected or created;
% 3. a general description of processing steps used;
% 4. any known imperfections or anomalies in the data.

%\clearpage

%Delete all unused file types below. Copy/paste for multiples of each file type as needed.
The supporting information presented here contains additional details in the computation of convective parameters, the interpretation of performance scores, an additional case study, and performance scores for the variable WMS.

\noindent\textbf{Text S1: Computation of convective parameters}

To compute the most unstable CAPE, the parcel profile is launched at the level of the highest equivalent potential temperature. WRF-python takes the surface temperature, pressure, and altitude as input, in addition to the pressure-level temperature, humidity, and geopotential height. Thus it accounts for the altitude of the surface for parcel profiles. CAPE itself is the integral between the parcel and environmental temperature profiles, where the parcel is buoyant.

\begin{equation}
    CAPE = g \cdot \int_{z_{LFC}}^{z_{LNB}} \frac{T'_{v} - T_{v}}{T_{v}} \,dz \, ,
\end{equation}

where T$_v$ is the virtual temperature of the environment, T'$_v$ the virtual temperature of the lifted parcel, g the gravity constant, z the altitude, LFC the level of free convection and LNB the level of neutral buoyancy.\\
To obtain the DLS, the magnitude of the differential vector between surface and mid-atmospheric winds is computed:

\begin{equation}
    DLS = \sqrt{(u_{500} - u_{sfc})^2 + (v_{500} - v_{sfc})^2} \, ,
\end{equation}

where u and v are the horizontal components of wind at the surface and 500 hPa level.\\
The compound parameter WMS combines DLS and CAPE in the following manner:

\begin{equation} \label{eq:wms}
    WMS = \sqrt{2 \cdot CAPE} \cdot DLS
\end{equation}

\noindent\textbf{Table S1: Model information}

Table \ref{tab:models} provides an overview of the available pressure levels for each model, their primary moisture variable, and the main training period.\\

\begin{table}[htbp]
    \centering
    \begin{tabular}{c|c|c|c}
        Model & Pressure levels &  Moisture   & Training   \\ 
        & & variable & period  \\ \hline
        IFS &  50,  100,  150,  200,  250,  300,  400,  500, & \\
        & 600,  700,  850,  925,
           1000 & Q\\
        ERA5 & 50,  100,  150,  200,  250,  300,  400,  500,  & \\
        &600,  700,  850,  925,
           1000 & Q\\
        Pangu-Weather* & 50,  100,  150,  200,  250,  300,  400,  500,  & \\
        &600,  700,  850,  925,
           1000 & Q & 1979-2017 \\
        GraphCast* & 1,    2,    3,    5,    7,   10,   20,   30,   50,   70,  100,  125, 150,  175,  200,  
                 & Q & 1979-2017 \\
            &  225,  250, 300,  350,  400, 450,  500,  550,  600,
            650,  700,  & \\
           &   750,  775, 800,  825,  850,  875,  900,  925,  950,  975,
           1000 & & \\
        GraphCast-oper* & 50,  100,  150,  200,  250,  300,  400,  500,  & \\
        &600,  700,  850,  925,
           1000 & Q & 1979-2017 \\
        FourCastNet* & 50,  100,  150,  200,  250,  300,  400,  500,  & \\
        &600,  700,  850,  925,
           1000 & RH & 1979-2015\\ \hline
         & Surface variables \\ \hline
        all datasets & 2m temperature, 10m wind, mean sea level pressure \\
        %ERA5 & 2m temperature, 10m wind, mean sea level pressure \\
        %Pangu-Weather* & 2m temperature, 10m wind, mean sea level pressure \\
        %GraphCast* & 2m temperature, 10m wind, mean sea level pressure \\
        %FourCastNet* & 2m temperature, 10m wind, mean sea level pressure \\
    \end{tabular}
    \caption{Model characteristics, neural weather models marked with an asterisk}
    \label{tab:models}
\end{table}
\FloatBarrier

\noindent\textbf{Table S2: Performance Scores}

Table \ref{tab:scores} summarizes the range of each forecast score, as well as its optimal value.\\
The FSS identifies the area fraction of the target variable exceeding a user-defined threshold. We use an area with a radius of 1° in which the fraction exceeding the threshold is evaluated. A FSS value of 1 indicates a perfect forecast (see Table \ref{tab:scores}). To interpret the SAL, a negative structure score indicates a too-fine spatial structure and too-small areas; a positive score indicates a too-large and too-flat distribution of values (smoothing). A low amplitude score indicates an underestimation, and a high score an overestimation. The location score accounts for both a mismatch in centroid, as well as a mismatch in overlap. A value of zero in each component indicates are perfect forecast (see Table \ref{tab:scores}).

\begin{table}[htbp]
    \centering
    \begin{tabular}{c|c|c}
        Score & Value range & Optimal score \\ \hline
        RMSE & $\geq$ 0 & 0 \\
        BIAS &  $\pm \infty$ & 0 \\
        FSS & 0,1 & 1 \\
        S & -2,2 & 0 \\
        A & -2,2 & 0 \\
        L & 0,2 & 0 \\
    \end{tabular}
    \caption{Forecast evaluation metrics}
    \label{tab:scores}
\end{table}
\FloatBarrier

\noindent\textbf{Table S3: Convective Regions}

Table \ref{tab:regions} indicates the area used for each region, as well as the months of their respective convective seasons.\\

\begin{table}[htbp]
    \centering
    \begin{tabular}{c|c|c|c}
        Region & Latitudes & Longitudes & Months  \\ \hline
        North America & 20/50 & 250/300 & April-August\\
        Europe & 35/55 & 350/30 & April-September\\
        Argentina & -40/-25 & 290/310 & September-February\\
        Australia & -35/-20 & 140/155 & September-February\\
%        China & 20/35 & 100/120 & April-September
    \end{tabular}
    \caption{Regions}
    \label{tab:regions}
\end{table}
\FloatBarrier

\noindent\textbf{Table S4: Forecast scores of April 12, 12 UTC, 12h lead-time}
The scores for both Tables \ref{tab:sc_short} and \ref{tab:sc_long} are computed including the sea area, and exclusively in the domain shown in Figs 1 and 2 of the main manuscript.

\begin{table}[htbp]
    \centering
    \begin{tabular}{c|c|c|c|c}
    %\singlespacing
        Score & IFS & Pangu-Weather-oper & GraphCast-oper & FourCastNet-oper \\ \hline
        \multicolumn{5}{c}{CAPE} \\ \hline
        RMSE & 312 & 286 &  \textbf{*271} &  386 \\
        BIAS & -88 & 90 & \textbf{*1} &  -92\\
        FSS$_{300}$ & 0.92 & 0.93 & \textbf{*0.96} & 0.93 \\
        FSS$_{1000}$ & 0.87 & \textbf{*0.91} & *0.88 & 0.61 \\
        S & 0.21 & 0.23 & \textbf{*0.02} & 0.27  \\
        A & 0.18 & 0.16 & \textbf{*-0.04} & -0.28  \\
        L & 0.03 & \textbf{*0.02} & 0.02 & 0.03\\ \hline
        \multicolumn{5}{c}{DLS} \\ \hline
        RMSE & 3.3 & 3.3 & \textbf{*2.9} & 3.5 \\
        BIAS & 0.2 & 0.2 & -0.2 & \textbf{*0.1}\\ \hline
        %FSS$_{10}$ & 0.99 & 0.99 & 0.99 & 0.99 \\
        %FSS$_{20}$ & 0.95 & 0.97 & 0.97 & 0.96 \\ \hline
        \multicolumn{5}{c}{WMS} \\ \hline
        RMSE & 332 & 303 & \textbf{*262} & 334 \\
        BIAS & 88 & 90 & \textbf{*-24} & -39 \\
        FSS$_{300}$ & 0.91 & 0.93 & \textbf{*0.96} & 0.92 \\
        FSS$_{500}$ & 0.91 & 0.93 & \textbf{*0.95} & 0.92 \\
        S & *0.31 & 0.29 & 0.11 & \textbf{0.03} \\
        A & 0.18 & 0.18 & \textbf{*-0.06} & -0.09 \\
        L & 0.03 & 0.05 & \textbf{*0.02} & *0.03 \\
    
    \end{tabular}
    \caption{Forecast scores at 12h lead-time, best model is marked in bold and with an asterisk}
    \label{tab:sc_short}
\end{table}
\FloatBarrier

\noindent\textbf{Table S5: Forecast scores of April 12, 12 UTC, 156h lead-time}

\begin{table}[htbp]
    \centering
    \begin{tabular}{c|c|c|c|c}
        Score & IFS & Pangu-Weather-oper & GraphCast-oper & FourCastNet-oper \\ \hline
        \multicolumn{5}{c}{CAPE} \\ \hline
        RMSE & 829 & 570 & \textbf{*520} & 606 \\
        BIAS & -223 & 210 & \textbf{*60} & -288 \\
        FSS$_{300}$ & 0.26 & 0.76 & \textbf{*0.77} & 0.51 \\
        FSS$_{1000}$ & 0.20 & 0.71 &\textbf{*0.74} & 0.36 \\
        S & 0.54 & 0.39 & \textbf{*0.17} & 0.63 \\
        A & -0.69 & 0.35 & \textbf{*0.09} & -1.06 \\
        L & 0.34 & 0.11 & \textbf{*0.09} & 0.36 \\ \hline
        \multicolumn{5}{c}{DLS} \\ \hline
        RMSE & 10.2 & \textbf{*5.1} & 5.1 & 7.7 \\
        BIAS & 3.2 & 1.2 & \textbf{*-0.2} & 1.2 \\ \hline
        %FSS$_{10}$ & 0.98 & 0.98 & 0.98 & 0.99 \\
        %FSS$_{20}$ & 0.83 & 0.83 & 0.89 & 0.92 \\ \hline
        \multicolumn{5}{c}{WMS} \\ \hline
        RMSE & 870 & 577 & \textbf{*511} & 604  \\
        BIAS & -124 & 216 & \textbf{*39} & -257 \\
        FSS$_{300}$ & 0.26 & 0.76 & \textbf{*0.77} & 0.51 \\
        FSS$_{500}$ & 0.26 & 0.71 & \textbf{*0.74} & 0.36 \\
        S & 0.33 & 0.51 & \textbf{*0.19} & 0.36 \\
        A & -0.33 & 0.40 & \textbf{*0.08} & -0.84 \\
        L & 0.35 & 0.11 & \textbf{*0.10} & 0.37 \\
    \end{tabular}
    \caption{Same as Table \ref{tab:sc_short} but for 156h lead-time}
    \label{tab:sc_long}
\end{table}
\FloatBarrier

\noindent\textbf{Figure S1: Comparison of vertical profiles for April 12, 2020, 12 UTC}

\begin{figure}[htbp]
    \centering
    \includegraphics[width=0.99\textwidth]{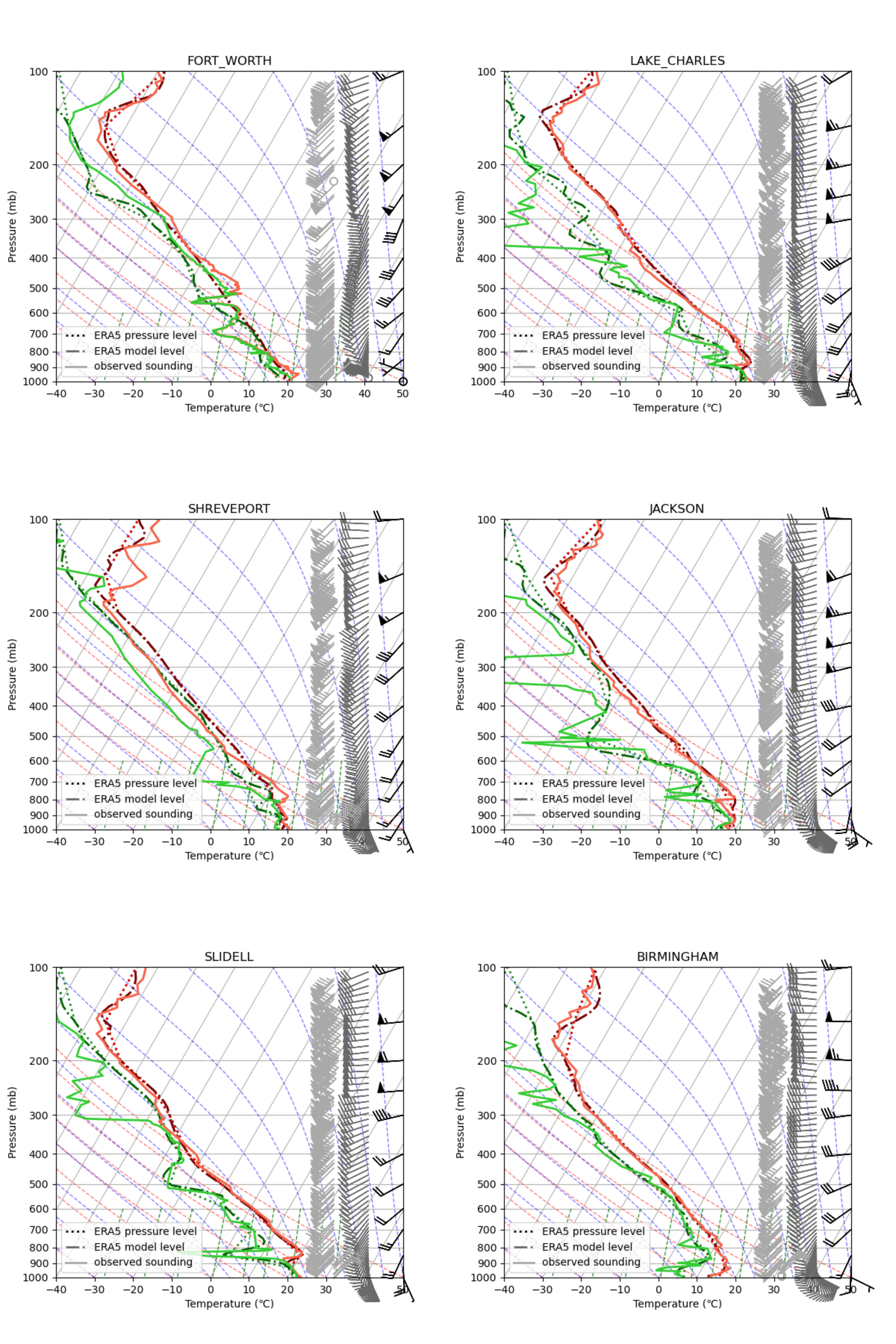}
    \caption{Comparison of soundings from ERA-5 pressure levels, model levels, and measured radiosoundings from 6 locations on April 12, 2020, 12 UTC}
    \label{fig:sound_comp}
\end{figure}
\FloatBarrier

\noindent\textbf{Figure S2: 12 UTC initializations for seasonal scores in the USA}

\begin{figure}[htbp]
    \centering
    \includegraphics[width=.75\textwidth]{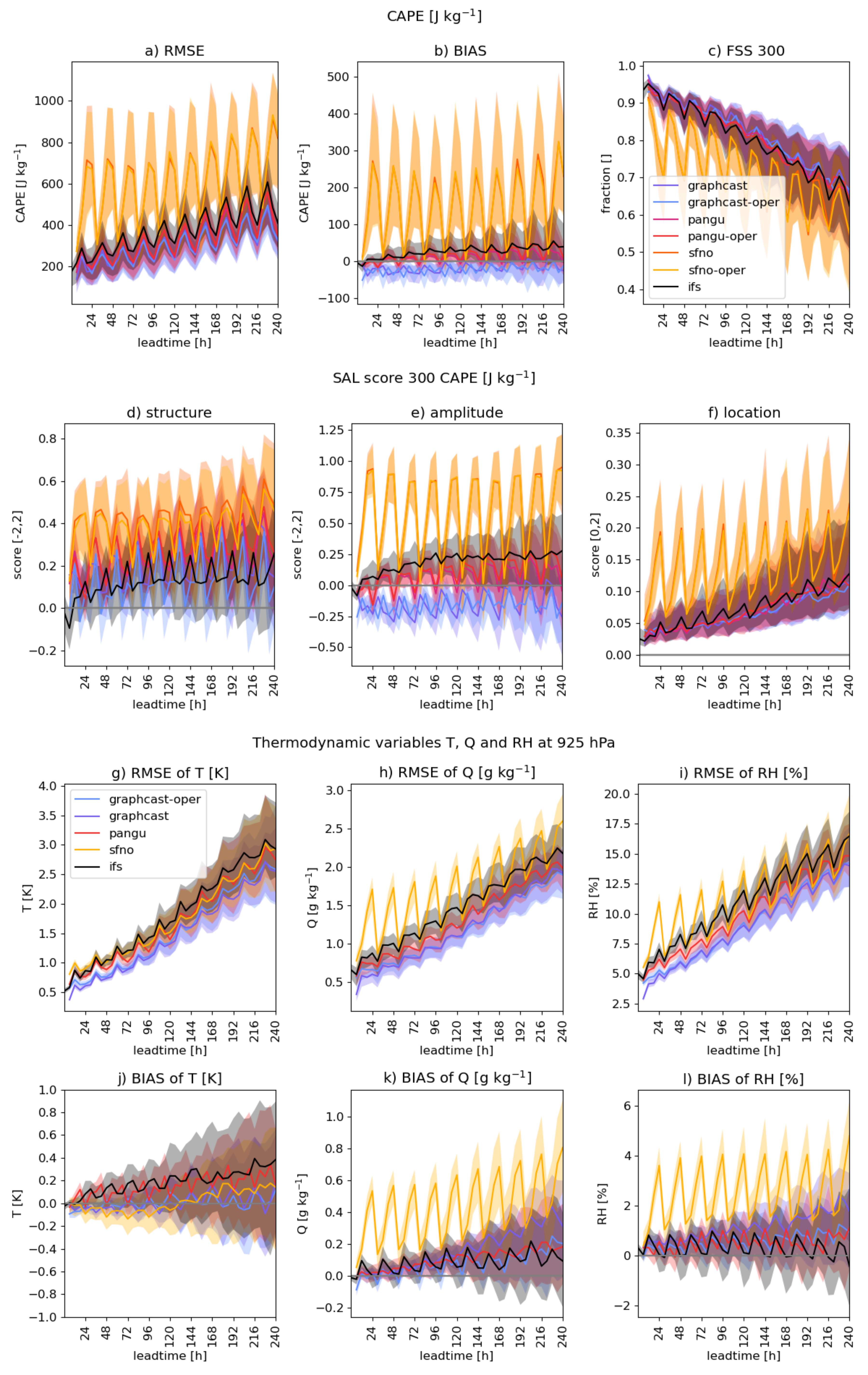}
    \caption{Seasonal evaluation of CAPE for North America at 12 UTC initialization shows Pangu-Weather and GraphCast performing better than IFS, with a moisture bias impacting the performance of FourCastNet. Top row: a) RMSE, b) BIAS, c) FSS$_{300}$; Second row: SAL components for CAPE \textgreater 300 J kg$^{-1}$; color legend in panel c); Third row: RMSE of T, Q and RH at 925 hPa (g-i); Fourth row:  BIAS (j-l); model legend in panel g). Solid line depicts the median and shading the interquartile range.}
    \label{fig:NA_scores_12}
\end{figure}

\FloatBarrier

\noindent\textbf{Figure S3: CAPE and DLS forecasts for July 7, 2020, 06 UTC}

\begin{figure}[htbp]
    \centering
    \includegraphics[width=0.99\textwidth]{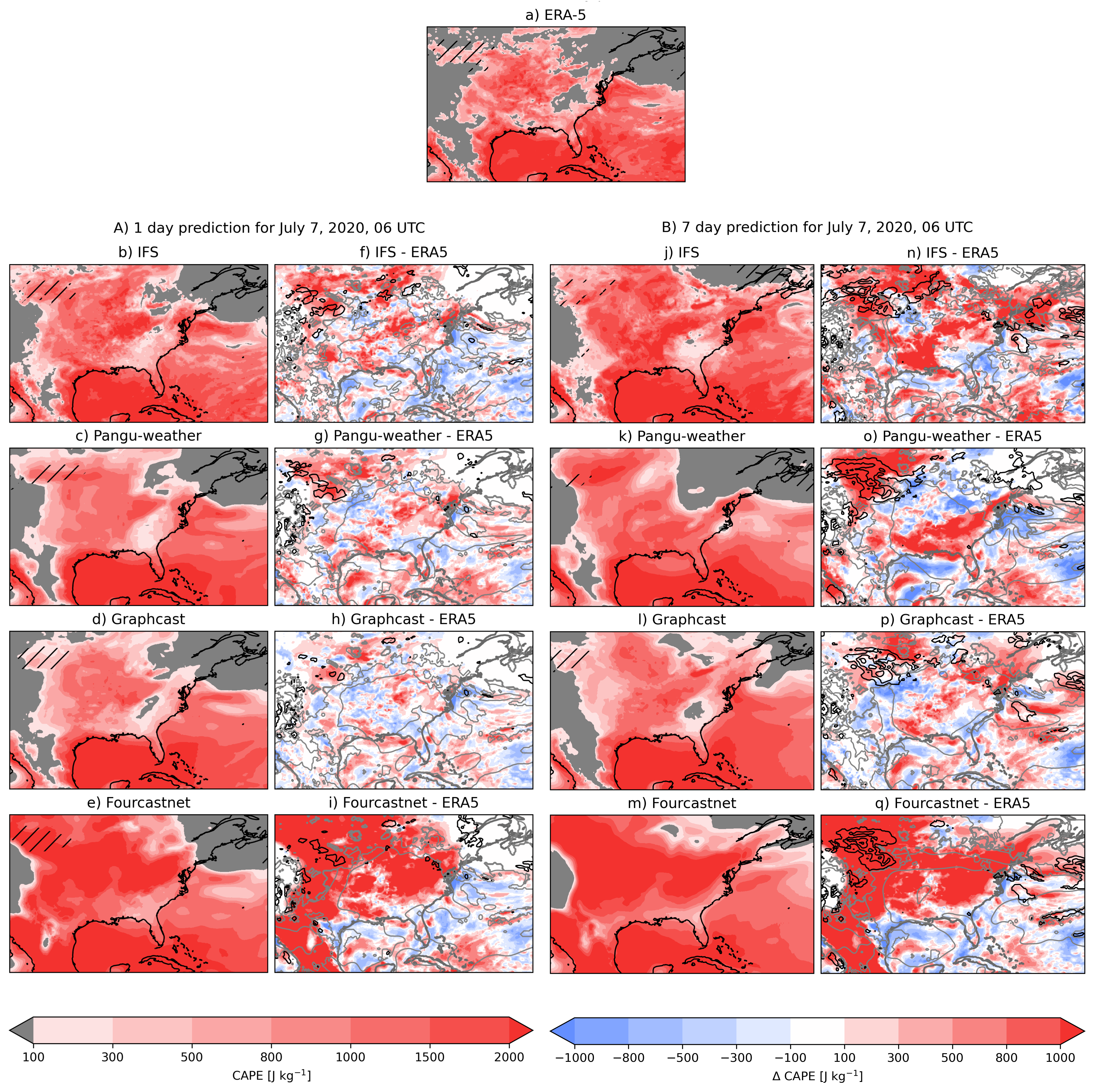}
    \caption{Comparison of forecasts of CAPE and DLS at 1 and 7 days lead-time; a) ERA-5 data of CAPE and DLS on July 7, 2020 at 06 UTC; A) 1 day forecast of CAPE and DLS (b-e) in comparison to ERA5 (f-i) B) 7 day forecast of CAPE and DLS (j-m) in comparison to ERA5 (n-q); hatched area indicates DLS \textgreater 20 m s$^{-1}$; contours indicated positive (grey) and negative (black) areas of $\Delta$ DLS in 5 m s$^{-1}$ increments}
    \label{fig:fc_comb_2}
\end{figure}
\FloatBarrier

\end{article}
\clearpage